\documentclass[showpacs]{revtex4}
\usepackage{indentfirst}
\begin{document}
\title{\bf Dibaryons with two heavy quarks}
\author{S.M. Gerasyuta}
\email{gerasyuta@SG6488.spb.edu}
\author{E.E. Matskevich}
\email{matskev@pobox.spbu.ru}
\affiliation{Department of Theoretical Physics, St. Petersburg State University, 198904,
St. Petersburg, Russia}
\affiliation{Department of Physics, LTA, 194021, St. Petersburg, Russia}
\begin{abstract}
The relativistic six-quark equations are constructed in the framework of the
dispersion relation technique. The relativistic six-quark amplitudes of dibaryons
including the light $u$, $d$ and heavy $c$, $b$ quarks are calculated.
The approximate solutions of these equations using the method based on the
extraction of leading singularities of the heavy hexaquark amplitudes are
obtained. The poles of these amplitudes determine the masses of charmed and
bottom dibaryons with the isospins $I=0$, $1$, $2$ and the spin-parities
$J^P=0^+$, $1^+$, $2^+$.
\end{abstract}
\maketitle
\section{Introduction.}
In 1977, Jaffe \cite{1} studied the color-magnetic interaction of the
one-gluon-exchange potential in the multiquark system and found that the
most attractive channel is the flavor singlet with quark content
$u^2d^2s^2$. The same symmetry analysis of the chiral boson exchange
potential leads to the similar result \cite{2}.

The $H$-particle, $N\Omega$-state and di-$\Omega$ may be strong interaction
stable. Up to now, these three interesting candidates of dibaryons are still
not found or confirmed by experiments. It seems that one should go beyond
these candidates and should search for the possible candidates in a wider region,
expecially the systems with heavy quarks, in terms of a more reliable model.

There were a number of theoretical predictions by using various models:
the quark cluster model \cite{3, 4}, the quark-delocation model \cite{5, 6}, the
chiral $SU(3)$ quark model \cite{7}, the flavor $SU(3)$ skyrmion model \cite{8}.
Lomon predicted a deuteronlike dibaryon resonance using R-matrix theory
\cite{9}. By employing the chiral $SU(3)$ quark model Zhang and Yu studied
$\Omega\Omega$ and $\Sigma\Omega$ states \cite{10, 11}. Lee and Yasui discuss
the stable multiquark states containing charm and bottom quark \cite{12}.

In a series of papers \cite{13, 14, 15, 16, 17} a method has been developed
which is convenient for analyzing relativistic three-hadron systems. The physics
of the three-hadron system can be described by means of a pair interaction
between the particles. There are three isobar channels, each of which
consists of a two-particle isobar and the third particle. The presence
of the isobar representation together with the condition of unitarity in
the pair energies and of analyticity leads to a system of integral equations
in a single variable. Their solution makes it possible to describe the
interaction of the produced particles in three-hadron systems.

In our papers \cite{18, 19, 20} relativistic generalization of the three-body
Faddeev equations was obtained in the form of dispersion relations in the
pair energy of two interacting quarks. The mass spectrum of $S$-wave
baryons including $u$, $d$, $s$ quarks was calculated by a method based on
isolating the leading singularities in the amplitude. We searched for the
approximate solution of integral three-quark equations by taking into
account two-particle and triangle singularities, and defined all
the smooth functions of the subenergy variables (as compared with the
singular part of the amplitude) in the middle point of the physical region
of Dalitz-plot, then the problem was reduced to the one of solving a system
of simple algebraic equations.

In the present paper the relativistic six-quark equations are found in
the framework of coupled-channel formalism. We use only planar diagrams; the other
diagrams due to the rules of $1/N_c$ expansion \cite{21, 22, 23} are neglected.
The six-quark amplitudes of dibaryons with two heavy quarks are calculated.
The poles of these amplitudes determine the masses of heavy dibaryons.

In Sec. II, the six-quark amplitudes of hexaquarks are constructed. The
dynamical mixing between the subamplitudes of dibaryons are considered.
The relativistic six-quark equations are constructed in the form of the
dispersion relation over the two-body subenergy. The approximate solutions
of these equations using the method based on the extraction of leading
singularities of the amplitude are obtained. Sec. III is devoted to the
calculation results for the dibaryon mass spectra (Tables \ref{tab1}, \ref{tab2}).
In conclusion, the status of the considered model is discussed.

\section{Six-quark amplitudes of the hexaquarks with the two heavy quarks.}

We derive the relativistic six-quark equations in the framework of the
dispersion relation technique. We use only planar diagrams; the other
diagrams due to the rules of $1/N_c$ expansion \cite{21, 22, 23} are neglected.
The current generates a six-quark system. The correct equations for the
amplitude are obtained by taking into account all possible subamplitudes.
Then one should represent a six-particle amplitude as a sum of 15 subamplitudes:

\begin{eqnarray}
A=\sum\limits_{i<j \atop i, j=1}^6 A_{ij}\, . \end{eqnarray}

This defines the division of the diagrams into groups according to the
certain pair interaction of particles. The total amplitude can be
represented graphically as a sum of diagrams. We need to consider only
one group of diagrams and the amplitude corresponding to them, for example
$A_{12}$. We shall consider the derivation of the relativistic generalization
of the Faddeev-Yakubovsky approach. In our case, the low-lying dibaryons
with the two heavy quarks are considered. We take into account the
pairwise interaction of all six quarks in the hexaquark.

For instance, we consider the state $\Sigma_c\Sigma_c$ with the isospin
$I=2$ and the spin-parity $J^P=0^+$ ($uuc\,\, uuc$). The set of diagrams
associated with the amplitude $A_{12}$ can further be broken down into
seven groups corresponding to subamplitudes:
$A_1^{1^{uu}}$, $A_1^{1^{cc}}$, $A_1^{0^{uc}}$, $A_2^{1^{uu}1^{uu}}$,
$A_2^{1^{uu}0^{uc}}$, $A_2^{0^{uc}0^{uc}}$, $A_3^{1^{uu}1^{uu}1^{cc}}$.

The system of graphical equations (see for example equation for the amplitude
$A_2^{0^{uc}0^{uc}}$ for the state $\Sigma_c\Sigma_c$ with the isospin
$I=2$ and the spin-parity $J^P=0^+$ ($uuc\,\, uuc$)) is determined by the
subamplitudes using the self-consistent method. The coefficients are determined
by the permutation of quarks.

In order to represent the subamplitudes $A_1^{1^{uu}}$, $A_1^{1^{cc}}$,
$A_1^{0^{uc}}$, $A_2^{1^{uu}1^{uu}}$, $A_2^{1^{uu}0^{uc}}$, $A_2^{0^{uc}0^{uc}}$,
$A_3^{1^{uu}1^{uu}1^{cc}}$ in the form of a dispersion relation, it is necessary
to define the amplitude of $qq$, $qQ$ and $QQ$ interactions. We use the results
of our relativistic quark model \cite{24} and write down the pair quark
amplitudes in the form:

\begin{equation}
a_n(s_{ik})=\frac{G^2_n(s_{ik})}
{1-B_n(s_{ik})} \, ,\end{equation}

\begin{equation}
B_n(s_{ik})=\int\limits_{(m_i+m_k)^2}^{\frac{(m_i+m_k)^2\Lambda}{4}}
\, \frac{ds'_{ik}}{\pi}\frac{\rho_n(s'_{ik})G^2_n(s'_{ik})}
{s'_{ik}-s_{ik}} \, ,\end{equation}

\begin{eqnarray}
\rho_n (s_{ik},J^P)&=&\left(\alpha(n,J^P) \frac{s_{ik}}{(m_i+m_k)^2}
+\beta(n,J^P)+\delta(n,J^P) \frac{(m_i-m_k)^2}{s_{ik}}\right)
\nonumber\\
&&\nonumber\\
&\times & \frac{\sqrt{(s_{ik}-(m_i+m_k)^2)(s_{ik}-(m_i-m_k)^2)}}
{s_{ik}}\, .
\end{eqnarray}

The coefficients $\alpha(n,J^P)$, $\beta(n,J^P)$ and
$\delta(n,J^P)$ are given in Table \ref{tab3}.
Here $n=1$ coresponds to $qq$, $qQ$ and $QQ$-pairs with $J^P=0^+$, $n=2$ corresponds
to $qq$, $qQ$ and $QQ$ with $J^P=1^+$.

The coupled integral equations correspond to Fig. 1 can be described
similar to \cite{25}.

Then we can go from the integration of the cosine
of the angles $dz_i$ to the integration over the subenergies.

Let us extract two- and three-particle singularities in the amplitudes
$A_1^{1^{uu}}$, $A_1^{1^{cc}}$, $A_1^{0^{uc}}$, $A_2^{1^{uu}1^{uu}}$,
$A_2^{1^{uu}0^{uc}}$, $A_2^{0^{uc}0^{uc}}$, $A_3^{1^{uu}1^{uu}1^{cc}}$:

\begin{eqnarray}
A_1^{1^{uu}}(s,s_{12345},s_{1234},s_{123},s_{12})&=
&\frac{\alpha_1^{1^{uu}} (s,s_{12345},s_{1234},s_{123},s_{12})
B_{1^{uu}}(s_{12})}{[1-B_{1^{uu}}(s_{12})]} \, ,\\
&&\nonumber\\
A_1^{1^{cc}}(s,s_{12345},s_{1234},s_{123},s_{12})&=
&\frac{\alpha_1^{1^{cc}} (s,s_{12345},s_{1234},s_{123},s_{12})
B_{1^{cc}}(s_{12})}{[1-B_{1^{cc}}(s_{12})]} \, ,\\
&&\nonumber\\
A_1^{0^{uc}}(s,s_{12345},s_{1234},s_{123},s_{12})&=
&\frac{\alpha_1^{0^{uc}} (s,s_{12345},s_{1234},s_{123},s_{12})
B_{0^{uc}}(s_{12})}{[1-B_{0^{uc}}(s_{12})]} \, ,\\
&&\nonumber\\
A_2^{1^{uu}1^{uu}}(s,s_{12345},s_{1234},s_{12},s_{34})&=
&\frac{\alpha_2^{1^{uu}1^{uu}} (s,s_{12345},s_{1234},s_{12},s_{34})
B_{1^{uu}}(s_{12})B_{1^{uu}}(s_{34})}{[1-B_{1^{uu}}(s_{12})]
[1-B_{1^{uu}}(s_{34})]} \, , \\
&&\nonumber\\
A_2^{1^{uu}0^{uc}}(s,s_{12345},s_{1234},s_{12},s_{34})&=
&\frac{\alpha_2^{1^{uu}0^{uc}} (s,s_{12345},s_{1234},s_{12},s_{34})
B_{1^{uu}}(s_{12})B_{0^{uc}}(s_{34})}{[1-B_{1^{uu}}(s_{12})]
[1-B_{0^{uc}}(s_{34})]} \, , \\
&&\nonumber\\
A_2^{0^{uc}0^{uc}}(s,s_{12345},s_{1234},s_{12},s_{34})&=
&\frac{\alpha_2^{0^{uc}0^{uc}} (s,s_{12345},s_{1234},s_{12},s_{34})
B_{0^{uc}}(s_{12})B_{0^{uc}}(s_{34})}{[1-B_{0^{uc}}(s_{12})]
[1-B_{0^{uc}}(s_{34})]} \, , \\
&&\nonumber\\
A_3^{1^{uu}1^{uu}1^{cc}}(s,s_{12345},s_{12},s_{34},s_{56})&=
&\frac{\alpha_3^{1^{uu}1^{uu}1^{cc}} (s,s_{12345},s_{12},s_{34},s_{56})
B_{1^{uu}}(s_{12})B_{1^{uu}}(s_{34}) B_{1^{cc}}(s_{56})}
{[1- B_{1^{uu}}(s_{12})] [1- B_{1^{uu}}(s_{34})][1- B_{1^{cc}}(s_{56})]}
 \, . \nonumber\\
&&
\end{eqnarray}

We used the classification of singularities, which was proposed in
paper \cite{26}. Using this classification, one defines the reduced
amplitudes $\alpha_1$, $\alpha_2$, $\alpha_3$ as well as the $B$-functions
in the middle point of physical region of Dalitz-plot at the point $s_0$.

Such choice of point $s_0$ allows us to replace integral equations
($\Sigma_c\Sigma_c$, $I=2$, $J^P=0^+$)
by the algebraic equations (\ref{12}) -- (\ref{18}):

\begin{eqnarray}
\label{12}
\alpha_1^{1^{uu}}&=&\lambda+4\, \alpha_1^{1^{uu}} I_1(1^{uu}1^{uu})
+4\, \alpha_1^{0^{uc}} I_1(1^{uu}0^{uc})
+2\, \alpha_2^{1^{uu}1^{uu}} I_2(1^{uu}1^{uu}1^{uu})
+8\, \alpha_2^{1^{uu}0^{uc}} I_2(1^{uu}1^{uu}0^{uc})
\nonumber\\
&&\nonumber\\
&+&2\, \alpha_2^{0^{uc}0^{uc}} I_2(1^{uu}0^{uc}0^{uc})\\
&&\nonumber\\
\label{13}
\alpha_1^{1^{cc}}&=&\lambda+8\, \alpha_1^{0^{uc}} I_1(1^{ss}0^{uc})
+12\, \alpha_2^{0^{uc}0^{uc}} I_2(1^{cc}0^{uc}0^{uc})\\
&&\nonumber\\
\label{14}
\alpha_1^{0^{uc}}&=&\lambda+3\, \alpha_1^{1^{uu}} I_1(0^{uc}1^{uu})
+\alpha_1^{1^{cc}} I_1(0^{uc}1^{cc})+4\, \alpha_1^{0^{uc}} I_1(0^{uc}0^{uc})
+6\, \alpha_2^{1^{uu}0^{uc}} I_2(0^{uc}1^{uu}0^{uc})
\nonumber\\
&&\nonumber\\
&+&3\, \alpha_2^{0^{uc}0^{uc}} I_2(0^{uc}0^{uc}0^{uc})\\
&&\nonumber\\
\label{15}
\alpha_2^{1^{uu}1^{uu}}&=&\lambda
+4\, \alpha_1^{1^{uu}} I_3(1^{uu}1^{uu}1^{uu})
+8\, \alpha_1^{0^{uc}} I_4(1^{uu}1^{uu}0^{uc})
+16\, \alpha_2^{1^{uu}0^{uc}} I_7(1^{uu}1^{uu}1^{uu}0^{uc})
\nonumber\\
&&\nonumber\\
&+&\alpha_2^{0^{uc}0^{uc}} (4\, I_5(1^{uu}1^{uu}0^{us}0^{uc})
+8\, I_6(1^{uu}1^{uu}0^{uc}0^{uc}))\\
&&\nonumber\\
\label{16}
\alpha_2^{1^{uu}0^{uc}}&=&\lambda
+\alpha_1^{1^{uu}} (2\, I_3(1^{uu}0^{uc}1^{uu})+I_4(0^{uc}1^{uu}1^{uu}))
+\alpha_1^{0^{uc}} (2\, I_3(1^{uu}0^{uc}0^{uc})+2\, I_4(1^{uu}0^{uc}0^{uc}))
\nonumber\\
&&\nonumber\\
&+&\alpha_2^{1^{uu}0^{uc}} (2\, I_5(1^{uu}0^{uc}1^{uu}0^{uc})
+2\, I_6(1^{uu}0^{us}0^{uc}1^{uu})+2\, I_7(1^{uu}0^{uc}1^{uu}0^{uc})
+2\, I_7(1^{uu}0^{us}0^{uc}1^{uu})
\nonumber\\
&&\nonumber\\
&+&2\, I_7(0^{uc}1^{uu}1^{uu}0^{uc}))
+\alpha_2^{0^{uc}0^{uc}} (I_5(0^{uc}1^{uu}0^{uc}0^{uc})
+2\, I_6(1^{uu}0^{uc}0^{uc}0^{uc})+2\, I_7(0^{uc}1^{uu}0^{uc}0^{uc}))
\nonumber\\
&&\\
\label{17}
\alpha_2^{0^{uc}0^{uc}}&=&\lambda+\alpha_1^{1^{uu}} (I_3(0^{uc}0^{uc}1^{uu})
+4\, I_4(0^{uc}0^{uc}1^{uu}))+\alpha_1^{1^{cc}} I_3(0^{uc}0^{uc}1^{cc})
+\alpha_1^{0^{uc}} (2\, I_3(0^{uc}0^{uc}0^{uc})
\nonumber\\
&&\nonumber\\
&+&4\, I_4(0^{uc}0^{uc}0^{uc}))
+2\, \alpha_2^{1^{uu}1^{uu}} I_6(0^{uc}0^{uc}1^{uu}1^{uu})
+\alpha_2^{1^{uu}0^{uc}} (4\, I_5(0^{uc}0^{uc}1^{uu}0^{uc})
\nonumber\\
&&\nonumber\\
&+&4\, I_6(0^{uc}0^{uc}1^{uu}0^{uc})+4\, I_7(0^{uc}0^{uc}1^{uu}0^{uc})
+4\, I_7(0^{uc}0^{uc}0^{uc}1^{uu}))
+\alpha_2^{0^{uc}0^{uc}} (2\, I_6(0^{uc}0^{uc}0^{uc}0^{uc})
\nonumber\\
&&\nonumber\\
&+&4\, I_7(0^{uc}0^{uc}0^{uc}0^{uc}))
+2\, \alpha_3^{1^{uu}1^{uu}1^{cc}} I_8(0^{uc}0^{uc}1^{uu}1^{cc}1^{uu})\\
&&\nonumber\\
\label{18}
\alpha_3^{1^{uu}1^{uu}1^{cc}}&=&\lambda
+4\, \alpha_1^{1^{uu}} I_9(1^{uu}1^{uu}1^{cc}1^{uu})
+8\, \alpha_1^{0^{uc}} I_9(1^{uu}1^{cc}1^{uu}0^{uc})
+16\, \alpha_2^{1^{uu}0^{uc}} I_{10}(1^{uu}1^{uu}1^{cc}1^{uu}0^{uc})
\nonumber\\
&&\nonumber\\
&+&8\, \alpha_2^{0^{uc}0^{uc}} I_{10}(1^{uu}1^{cc}1^{uu}0^{uc}0^{uc})\, ,
\end{eqnarray}

\noindent
where $\lambda_i$ are the current constants. We used the functions
$I_1$, $I_2$, $I_3$, $I_4$, $I_5$, $I_6$, $I_7$, $I_8$, $I_9$, $I_{10}$
similar to the paper \cite{27}.

The solutions of the system of equations are considered as:

\begin{equation}
\alpha_i(s)=\frac{F_i(s,\lambda_i)}{D(s)} \, ,\end{equation}

\noindent
where zeros of $D(s)$ determinants define the masses of bound states of
dibaryons.

\section{Calculation results.}

The model in question take into account the hexaquarks with the two
heavy quarks $qqqqQQ$, $q=u, d$, $Q=c, b$: $uuuucc$, $uuudcc$, $uuddcc$,
$uuuubb$, $uuudbb$, $uuddbb$.

The quark masses of the model are $m_q=495\, MeV$, $m_c=1655\, MeV$ and
$m_b=4840\, MeV$.

The experimental data are absent, therefore we use the dimensionless
parameters, which are similar to the previous paper \cite{25}. It allows
us to calculate the mass spectra of $qqqqQQ$ states. We use the gluon
coupling constants $g_0=0.653$ (diquark $J^P=0^+$) and $g_1=0.292$
(diquark $J^P=1^+$), cutoff parameter $\Lambda=11$. We consider the
$\Lambda_{qc, cc}=8.52$, which are determined by $M=5250\, MeV$ ($IJ=22$
$\Sigma_c \Sigma^*_c$, $\Sigma^*_c \Sigma^*_c$), the threshold is $5290\, MeV$.
In the case of $b$-quarks the $\Lambda_{qb, bb}=7.35$ is determined by $M=11620\, MeV$
($IJ=22$ $\Sigma_b \Sigma^*_b$, $\Sigma^*_b \Sigma^*_b$), the threshold is
$11660\, MeV$.

We have calculated the heavy dibaryon masses with isospin $I=0$, $1$, $2$ and
spin-parity $J^P=0^+$, $1^+$, $2^+$, which are given in the Tables \ref{tab1}
and \ref{tab2}. The relativistic six-body approach possesses the dynamical
mixing and allows us to calculate the contributions of the subamplitudes to
the hexaquark amplitude (Tables \ref{tab4} -- \ref{tab6}). The calculated
dibaryon subamplitudes $A_2$ present the main contributions to the
hexaquark amplitude (about 70 percents). We use only two new parameters
for the calculation of $23$ $qqqqcc$ states and $19$ $qqqqbb$ states.

The lowest mass for the $qqqqcc$ states is $M=4364\, MeV$, for the $qqqqbb$
states is $M=8670\, MeV$.

In quark models, which describe rather well the masses and static properties
of hadrons, the masses of the quarks usually have the similar values for the
spectra of light and heavy hadrons. However, this is achieved at the expence
of some difference in the characteristic of the confinement potential. It
should be borne in mind that for a fixed hadron mass the masses of the
constituent quarks which enter into the composition of the hadron will become
smaller when the slope of the confinement potential increases or its radius
decreases. Therefore, conversely, we can change the masses of the constituent
quarks when going from the spectrum of light to the heavy hadrons, while
keeping the characteristic of the confinement potential unchanged. We can
effectively take into account the contribution of the confinement potential
in obtaining the spectrum of heavy hadrons. We neglect with the mass
distinction of $u$ and $d$ quarks. The estimation of the theoretical error
on the heavy dibaryons masses is $1\, MeV$. This result was obtained by the
choice of model parameters.

\section{Conclusions.}

In a strongly bound systems, which include the light quarks, where
$p/m \sim 1$, the approximation of nonrelativistic kinematics and dynamics
is not justified. In our paper, the relativistic description of six-particles
amplitudes of heavy dibaryons with the two heavy quarks is considered. We take
into account the $u$, $d$, $c$, $b$ quarks. Our model is confined to the
quark-antiquark pair production on account of the phase space restriction.
Here $m_q$ and $m_Q$ the ''mass'' of the constituent quark. Therefore the
production of new quark-antiquark pair is absent for the low-lying hadrons.

Hadronic molecules are loosely states of hadrons, whose inter-hadron distances
are larger than the quark confinement size.

The heavy analogue of $H$ dibaryon ($\Lambda_c\Lambda_c$) does not exist
though its potential is attractive \cite{28}. Oka et al. believe that the
future studies may specify the binding energy of such a molecule state. The
binding energy is sensitive to the cutoff parameter \cite{29}.
Our calculation allows us to obtain the $\Lambda_c\Lambda_c$ molecule
bound state (Table \ref{tab1}). There exist a loosely bound state with a
small binding energy $E_B=54\, MeV$. In the case of $\Sigma_c\Sigma_c$
dibaryon we obtain the bound state, but the binding energy is equal to
$E_B=540\, MeV$.

The similar results are obtained for the other heavy dibaryons with the
two heavy quarks (Tables \ref{tab1} -- \ref{tab2}). For the $\Lambda_b\Lambda_b$
system does not exist a loosely bound state with a small binding energy.
The heavy dibaryons and heavy baryon-antibaryons may be produced at LHC.

\begin{acknowledgments}
Gerasyuta S.M. would like to thank T. Barnes for useful discussions.
The work was carried with the support of the Russian Ministry of Education
(grant 2.1.1.68.26).
\end{acknowledgments}

\begin{table}
\caption{S-wave charmed dibaryon masses. Parameters of model: cutoff
$\Lambda=11.0$ and $\Lambda_{qc, cc}=8.52$, gluon coupling constants
$g_0=0.653$ and $g_1=0.292$. Quark masses $m_q=495\, MeV$ and
$m_c=1655\, MeV$.}
\label{tab1}
\begin{tabular}{|c|c|c|c|}
\hline
$I$ & $J$ & Dibaryons (quark content) & Mass (MeV) \\
\hline
2 & 0 & $\Sigma_c\Sigma_c$, $\Sigma^*_c\Sigma^*_c$ ($uuc\,\, uuc$) & 4933 \\
  &   & $\Delta\Xi^*_{cc}$ ($uuu\,\, ucc$)                         & 5231 \\
  & 1 & $\Sigma_c\Sigma_c$, $\Sigma_c\Sigma^*_c$,
                            $\Sigma^*_c\Sigma^*_c$ ($uuc\,\, uuc$) & 4933 \\
  &   & $\Delta\Xi_{cc}$, $\Delta\Xi^*_{cc}$ ($uuu\,\, ucc$)       & 5231 \\
  & 2 & $\Sigma_c\Sigma^*_c$, $\Sigma^*_c\Sigma^*_c$ ($uuc\,\, uuc$)
                                                                   & 5250 \\
  &   & $\Delta\Xi_{cc}$, $\Delta\Xi^*_{cc}$ ($uuu\,\, ucc$)       & 5231 \\
1 & 0 & $\Sigma_c\Sigma_c$, $\Sigma^*_c\Sigma^*_c$, $\Sigma_c\Lambda_c$
                                                   ($uuc\,\, udc$) & 4420 \\
  &   & $\Delta\Xi^*_{cc}$ ($uuu\,\, dcc+uud\,\, ucc$)             & 4956 \\
  &   & $N\Xi_{cc}$ ($uud\,\, ucc$)                                & 4956 \\
  & 1 & $\Sigma_c\Sigma_c$, $\Sigma_c\Sigma^*_c$, $\Sigma^*_c\Sigma^*_c$,
        $\Sigma_c\Lambda_c$, $\Sigma^*_c\Lambda_c$ ($uuc\,\, udc$) & 4420 \\
  &   & $\Delta\Xi_{cc}$, $\Delta\Xi^*_{cc}$ ($uuu\,\, dcc+uud\,\, ucc$)
                                                                   & 4956 \\
  &   & $N\Xi_{cc}$, $N\Xi^*_{cc}$ ($uud\,\, ucc$)                 & 4956 \\
  & 2 & $\Sigma_c\Sigma^*_c$, $\Sigma^*_c\Sigma^*_c$,
                             $\Sigma^*_c\Lambda_c$ ($uuc\,\, udc$) & 4911 \\
  &   & $\Delta\Xi_{cc}$, $\Delta\Xi^*_{cc}$ ($uuu\,\, dcc+uud\,\, ucc$)
                                                                   & 4999 \\
  &   & $N\Xi^*_{cc}$ ($uud\,\, ucc$)                              & 5136 \\
0 & 0 & $\Sigma_c\Sigma_c$, $\Sigma^*_c\Sigma^*_c$,
                                       ($uuc\,\, ddc+udc\,\, udc$) & 4364 \\
  &   & $\Sigma_c\Lambda_c$, $\Lambda_c\Lambda_c$ ($udc\,\, udc$)  & 4516 \\
  &   & $N\Xi_{cc}$, $\Delta\Xi^*_{cc}$ ($uud\,\, dcc+udd\,\, ucc$)
                                                                   & 4740 \\
  & 1 & $\Sigma_c\Sigma_c$, $\Sigma_c\Sigma^*_c$, $\Sigma^*_c\Sigma^*_c$,
                                       ($uuc\,\, ddc+udc\,\, udc$) & 4364 \\
  &   & $\Sigma_c\Lambda_c$, $\Sigma^*_c\Lambda_c$ $\Lambda_c\Lambda_c$
                                                  ($udc\,\, udc$)  & 4516 \\
  &   & $N\Xi_{cc}$, $N\Xi^*_{cc}$, $\Delta\Xi_{cc}$, $\Delta\Xi^*_{cc}$
                                       ($uud\,\, dcc+udd\,\, ucc$) & 4740 \\
  & 2 & $\Sigma_c\Sigma^*_c$, $\Sigma^*_c\Sigma^*_c$,
                                       ($uuc\,\, ddc+udc\,\, udc$) & 5086 \\
  &   & $N\Xi^*_{cc}$, $\Delta\Xi_{cc}$, $\Delta\Xi^*_{cc}$
                                       ($uud\,\, dcc+udd\,\, ucc$) & 5029 \\
\hline
\end{tabular}
\end{table}

\begin{table}
\caption{S-wave bottom dibaryon masses. Parameters of model: cutoff
$\Lambda=11.0$ and $\Lambda_{qb, bb}=7.35$, gluon coupling constants
$g_0=0.653$ and $g_1=0.292$. Quark masses $m_q=495\, MeV$ and
$m_b=4840\, MeV$.}
\label{tab2}
\begin{tabular}{|c|c|c|c|}
\hline
$I$ & $J$ & Dibaryons (quark content) & Mass (MeV) \\
\hline
2 & 0 & $\Sigma_b\Sigma_b$, $\Sigma^*_b\Sigma^*_b$ ($uub\,\, uub$) & 10290 \\
  &   & $\Delta\Xi^*_{bb}$ ($uuu\,\, ubb$)                         & -- \\
  & 1 & $\Sigma_b\Sigma_b$, $\Sigma_b\Sigma^*_b$,
                            $\Sigma^*_b\Sigma^*_b$ ($uub\,\, uub$) & 10290 \\
  &   & $\Delta\Xi_{bb}$, $\Delta\Xi^*_{bb}$ ($uuu\,\, ubb$)       & -- \\
  & 2 & $\Sigma_b\Sigma^*_b$, $\Sigma^*_b\Sigma^*_b$ ($uub\,\, uub$)
                                                                   & 11620 \\
  &   & $\Delta\Xi_{bb}$, $\Delta\Xi^*_{bb}$ ($uuu\,\, ubb$)       & -- \\
1 & 0 & $\Sigma_b\Sigma_b$, $\Sigma^*_b\Sigma^*_b$, $\Sigma_b\Lambda_b$
                                                   ($uub\,\, udb$) & 8670 \\
  &   & $\Delta\Xi^*_{bb}$ ($uuu\,\, dbb+uud\,\, ubb$)             & 11395 \\
  &   & $N\Xi_{bb}$ ($uud\,\, ubb$)                                & 11395 \\
  & 1 & $\Sigma_b\Sigma_b$, $\Sigma_b\Sigma^*_b$, $\Sigma^*_b\Sigma^*_b$,
        $\Sigma_b\Lambda_b$, $\Sigma^*_b\Lambda_b$ ($uub\,\, udb$) & 8670 \\
  &   & $\Delta\Xi_{b}$, $\Delta\Xi^*_{bb}$ ($uuu\,\, dbb+uud\,\, ubb$)
                                                                   & 11395 \\
  &   & $N\Xi_{bb}$, $N\Xi^*_{b}$ ($uud\,\, ubb$)                 & 11395 \\
  & 2 & $\Sigma_b\Sigma^*_b$, $\Sigma^*_b\Sigma^*_b$,
                             $\Sigma^*_b\Lambda_b$ ($uub\,\, udb$) & 10715 \\
  &   & $\Delta\Xi_{bb}$, $\Delta\Xi^*_{bb}$ ($uuu\,\, dbb+uud\,\, ubb$)
                                                                   & 11372 \\
  &   & $N\Xi^*_{bb}$ ($uud\,\, ubb$)                              & -- \\
0 & 0 & $\Sigma_b\Sigma_b$, $\Sigma^*_b\Sigma^*_b$,
                                       ($uub\,\, ddb+udb\,\, udb$) & 8482 \\
  &   & $\Sigma_b\Lambda_b$, $\Lambda_b\Lambda_b$ ($udb\,\, udb$)  & 9175 \\
  &   & $N\Xi_{bb}$, $\Delta\Xi^*_{bb}$ ($uud\,\, dbb+udd\,\, ubb$)
                                                                   & 10828 \\
  & 1 & $\Sigma_b\Sigma_b$, $\Sigma_b\Sigma^*_b$, $\Sigma^*_b\Sigma^*_b$,
                                       ($uub\,\, ddb+udb\,\, udb$) & 8482 \\
  &   & $\Sigma_b\Lambda_b$, $\Sigma^*_b\Lambda_b$ $\Lambda_b\Lambda_b$
                                                  ($udb\,\, udb$)  & 9175 \\
  &   & $N\Xi_{bb}$, $N\Xi^*_{bb}$, $\Delta\Xi_{bb}$, $\Delta\Xi^*_{bb}$
                                       ($uud\,\, dbb+udd\,\, ubb$) & 10828 \\
  & 2 & $\Sigma_b\Sigma^*_b$, $\Sigma^*_b\Sigma^*_b$,
                                       ($uub\,\, ddb+udb\,\, udb$) & 11518 \\
  &   & $N\Xi^*_{bb}$, $\Delta\Xi_{bb}$, $\Delta\Xi^*_{bb}$
                                       ($uud\,\, dbb+udd\,\, ubb$) & 11583 \\
\hline
\end{tabular}
\end{table}

\begin{table}
\caption{Vertex functions and Chew-Mandelstam coefficients.}\label{tab3}
\begin{tabular}{|c|c|c|c|c|}
\hline
$i$ & $G_i^2(s_{kl})$ & $\alpha_i$ & $\beta_i$ & $\delta_i$ \\
\hline
& & & & \\
$0^+$ & $\frac{4g}{3}-\frac{8gm_{kl}^2}{(3s_{kl})}$
& $\frac{1}{2}$ & $-\frac{1}{2}\frac{(m_k-m_l)^2}{(m_k+m_l)^2}$ & $0$ \\
& & & & \\
$1^+$ & $\frac{2g}{3}$ & $\frac{1}{3}$
& $\frac{4m_k m_l}{3(m_k+m_l)^2}-\frac{1}{6}$
& $-\frac{1}{6}\frac{(m_k-m_l)^2}{(m_k+m_l)^2}$ \\
& & & & \\
\hline
\end{tabular}
\end{table}

\begin{table}
\caption{$IJ=00$ $\Sigma_Q\Sigma_Q$, $\Sigma^*_Q\Sigma^*_Q$.
Parameters of model: cutoff $\Lambda=11.0$, $\Lambda_{qc, cc}=8.52$,
$\Lambda_{qb, bb}=7.35$, gluon coupling constants
$g_0=0.653$ and $g_1=0.292$. Quark masses $m_q=495\, MeV$,
$m_c=1655\, MeV$ and $m_b=4840\, MeV$.}\label{tab4}
\begin{tabular}{|c|c|c|}
\hline
 Subamplitudes &
\multicolumn{2}{|c|}{Contributions, percent}\\
\cline{2-3} & $Q=c$ & $Q=b$ \\
\hline
$A_1^{1^{uu}}$ & 2.6 & 4.6 \\
$A_1^{1^{dd}}$ & 2.6 & 4.6 \\
$A_1^{1^{QQ}}$ & 3.7 & 6.4 \\
$A_1^{0^{ud}}$ & 6.4 & 10.2 \\
$A_1^{0^{uQ}}$ & 3.1 & 2.5 \\
$A_1^{0^{dQ}}$ & 3.1 & 2.5 \\
$A_3^{1^{uu}1^{dd}1^{QQ}}$ & 2.6 & 1.8 \\
$A_3^{0^{ud}0^{uQ}0^{dQ}}$ & 2.0 & 0.2 \\
$A_2^{1^{uu}1^{dd}}$ & 6.1 & 10.2 \\
$A_2^{1^{uu}0^{dQ}}$ & 2.8 & 2.2 \\
$A_2^{1^{dd}0^{uQ}}$ & 2.8 & 2.2 \\
$A_2^{0^{ud}0^{ud}}$ & 25.4 & 35.5 \\
$A_2^{0^{ud}0^{uQ}}$ & 7.4 & 5.0 \\
$A_2^{0^{ud}0^{dQ}}$ & 7.4 & 5.0 \\
$A_2^{0^{uQ}0^{uQ}}$ & 7.8 & 2.6 \\
$A_2^{0^{dQ}0^{dQ}}$ & 7.8 & 2.6 \\
$A_2^{0^{uQ}0^{dQ}}$ & 6.6 & 2.2 \\
\hline
$\sum A_1$ & 21.4 & 30.6 \\
$\sum A_2$ & 74.0 & 67.5 \\
$\sum A_3$ & 4.6 & 2.0 \\
\hline
\end{tabular}
\end{table}

\begin{table}
\caption{$IJ=00$ $\Sigma_Q\Lambda_Q$, $\Lambda_Q\Lambda_Q$.
Parameters of model: cutoff $\Lambda=11.0$, $\Lambda_{qc, cc}=8.52$,
$\Lambda_{qb, bb}=7.35$, gluon coupling constants
$g_0=0.653$ and $g_1=0.292$. Quark masses $m_q=495\, MeV$,
$m_c=1655\, MeV$ and $m_b=4840\, MeV$.}\label{tab5}
\begin{tabular}{|c|c|c|}
\hline
 Subamplitudes &
\multicolumn{2}{|c|}{Contributions, percent}\\
\cline{2-3} & $Q=c$ & $Q=b$ \\
\hline
$A_1^{1^{uu}}$ & 3.0 & 5.6 \\
$A_1^{1^{dd}}$ & 3.0 & 5.6 \\
$A_1^{1^{QQ}}$ & 4.3 & 8.3 \\
$A_1^{0^{ud}}$ & 7.0 & 11.4 \\
$A_1^{0^{uQ}}$ & 3.6 & 3.2 \\
$A_1^{0^{dQ}}$ & 3.6 & 3.2 \\
$A_3^{0^{ud}0^{uQ}0^{dQ}}$ & 3.1 & 0.3 \\
$A_2^{0^{ud}0^{ud}}$ & 28.8 & 39.8 \\
$A_2^{0^{ud}0^{uQ}}$ & 9.2 & 6.9 \\
$A_2^{0^{ud}0^{dQ}}$ & 9.2 & 6.9 \\
$A_2^{0^{uQ}0^{uQ}}$ & 10.1 & 3.6 \\
$A_2^{0^{dQ}0^{dQ}}$ & 10.1 & 3.6 \\
$A_2^{0^{uQ}0^{dQ}}$ & 5.1 & 1.8 \\
\hline
$\sum A_1$ & 24.5 & 37.2 \\
$\sum A_2$ & 72.4 & 62.5 \\
$\sum A_3$ & 3.1 & 0.3 \\
\hline
\end{tabular}
\end{table}

\begin{table}
\caption{$IJ=00$ $N\Xi_{QQ}$, $\Delta\Xi^*_{QQ}$.
Parameters of model: cutoff $\Lambda=11.0$, $\Lambda_{qc, cc}=8.52$,
$\Lambda_{qb, bb}=7.35$, gluon coupling constants
$g_0=0.653$ and $g_1=0.292$. Quark masses $m_q=495\, MeV$,
$m_c=1655\, MeV$ and $m_b=4840\, MeV$.}\label{tab6}
\begin{tabular}{|c|c|c|}
\hline
 Subamplitudes &
\multicolumn{2}{|c|}{Contributions, percent}\\
\cline{2-3} & $Q=c$ & $Q=b$ \\
\hline
$A_1^{1^{uu}}$ & 2.1 & 1.8 \\
$A_1^{1^{dd}}$ & 2.1 & 1.8 \\
$A_1^{1^{QQ}}$ & 3.6 & 5.7 \\
$A_1^{0^{ud}}$ & 6.6 & 6.2 \\
$A_1^{0^{uQ}}$ & 3.7 & 2.1 \\
$A_1^{0^{dQ}}$ & 3.7 & 2.1 \\
$A_3^{1^{uu}1^{dd}1^{QQ}}$ & 4.9 & 4.8 \\
$A_3^{0^{ud}0^{uQ}0^{dQ}}$ & 4.9 & 0.6 \\
$A_2^{1^{uu}1^{QQ}}$ & 10.6 & 15.7 \\
$A_2^{1^{dd}1^{QQ}}$ & 10.6 & 15.7 \\
$A_2^{1^{uu}0^{dQ}}$ & 3.0 & 10.0 \\
$A_2^{1^{dd}0^{uQ}}$ & 3.0 & 10.0 \\
$A_2^{1^{QQ}0^{ud}}$ & 22.3 & 34.6 \\
$A_2^{0^{ud}0^{uQ}}$ & 9.5 & 3.6 \\
$A_2^{0^{ud}0^{dQ}}$ & 9.5 & 3.6 \\
\hline
$\sum A_1$ & 21.7 & 19.6 \\
$\sum A_2$ & 68.6 & 75.1 \\
$\sum A_3$ & 9.8 & 5.4 \\
\hline
\end{tabular}
\end{table}

\newpage

\begin{picture}(600,600)
\put(0,545){\line(1,0){18}}
\put(0,547){\line(1,0){17.5}}
\put(0,549){\line(1,0){17}}
\put(0,551){\line(1,0){17}}
\put(0,553){\line(1,0){17.5}}
\put(0,555){\line(1,0){18}}
\put(30,550){\circle{25}}
\put(19,546){\line(1,1){15}}
\put(22,541){\line(1,1){17}}
\put(27.5,538.5){\line(1,1){14}}
\put(31,563){\vector(1,1){20}}
\put(31,538){\vector(1,-1){20}}
\put(47.5,560){\circle{16}}
\put(47.5,540){\circle{16}}
\put(55,564){\vector(3,2){18}}
\put(55,536){\vector(3,-2){18}}
\put(55,564){\vector(3,-2){18}}
\put(55,536){\vector(3,2){18}}
\put(78,575){1}
\put(78,553){2}
\put(78,541){3}
\put(78,518){4}
\put(54,580){5}
\put(54,513){6}
\put(40.5,556){\small $0^{uc}$}
\put(40.5,536){\small $0^{uc}$}
\put(90,548){$=$}
\put(110,545){\line(1,0){19}}
\put(110,547){\line(1,0){21}}
\put(110,549){\line(1,0){23}}
\put(110,551){\line(1,0){23}}
\put(110,553){\line(1,0){21}}
\put(110,555){\line(1,0){19}}
\put(140,560){\circle{16}}
\put(140,540){\circle{16}}
\put(147.5,564){\vector(3,2){18}}
\put(147.5,536){\vector(3,-2){18}}
\put(147.5,564){\vector(3,-2){18}}
\put(147.5,536){\vector(3,2){18}}
\put(128,555){\vector(1,3){11}}
\put(128,545){\vector(1,-3){11}}
\put(170,575){1}
\put(170,553){2}
\put(170,541){3}
\put(170,518){4}
\put(143,586){5}
\put(143,508){6}
\put(133,557){\small $0^{uc}$}
\put(133,537){\small $0^{uc}$}
\put(190,548){$+$}
\put(212,545){\line(1,0){18}}
\put(212,547){\line(1,0){17.5}}
\put(212,549){\line(1,0){17}}
\put(212,551){\line(1,0){17}}
\put(212,553){\line(1,0){17.5}}
\put(212,555){\line(1,0){18}}
\put(242,550){\circle{25}}
\put(231,546){\line(1,1){15}}
\put(234,541){\line(1,1){17}}
\put(239.5,538.5){\line(1,1){14}}
\put(243,563){\vector(1,1){20}}
\put(243,538){\vector(1,-1){20}}
\put(266,580){5}
\put(266,513){6}
\put(262,550){\circle{16}}
\put(270,550){\vector(3,2){17}}
\put(270,550){\vector(3,-2){17}}
\put(247,561){\vector(1,0){40}}
\put(247,539){\vector(1,0){40}}
\put(295,560){\circle{16}}
\put(295,540){\circle{16}}
\put(303,561){\vector(3,1){20}}
\put(303,561){\vector(3,-1){20}}
\put(303,539){\vector(3,1){20}}
\put(303,539){\vector(3,-1){20}}
\put(328,570){1}
\put(328,553){2}
\put(328,541){3}
\put(328,524){4}
\put(270,566){1}
\put(270,554){\small 2}
\put(270,540){\small 3}
\put(270,528){4}
\put(255,547){\small $1^{uu}$}
\put(288,557){\small $0^{uc}$}
\put(288,537){\small $0^{uc}$}
\put(346,548){$+$}
\put(368,545){\line(1,0){18}}
\put(368,547){\line(1,0){17.5}}
\put(368,549){\line(1,0){17}}
\put(368,551){\line(1,0){17}}
\put(368,553){\line(1,0){17.5}}
\put(368,555){\line(1,0){18}}
\put(398,550){\circle{25}}
\put(387,546){\line(1,1){15}}
\put(390,541){\line(1,1){17}}
\put(395.5,538.5){\line(1,1){14}}
\put(399,563){\vector(1,1){20}}
\put(399,538){\vector(1,-1){20}}
\put(422,580){5}
\put(422,513){6}
\put(418,550){\circle{16}}
\put(426,550){\vector(3,2){17}}
\put(426,550){\vector(3,-2){17}}
\put(403,561){\vector(1,0){40}}
\put(403,539){\vector(1,0){40}}
\put(451,560){\circle{16}}
\put(451,540){\circle{16}}
\put(459,561){\vector(3,1){20}}
\put(459,561){\vector(3,-1){20}}
\put(459,539){\vector(3,1){20}}
\put(459,539){\vector(3,-1){20}}
\put(482,570){1}
\put(482,553){2}
\put(482,541){3}
\put(482,524){4}
\put(426,566){1}
\put(426,554){\small 2}
\put(426,540){\small 3}
\put(426,528){4}
\put(411,547){\small $1^{cc}$}
\put(444,557){\small $0^{uc}$}
\put(444,537){\small $0^{uc}$}
\put(10,448){$+$}
\put(24,448){2}
\put(37,445){\line(1,0){18}}
\put(37,447){\line(1,0){17.5}}
\put(37,449){\line(1,0){17}}
\put(37,451){\line(1,0){17}}
\put(37,453){\line(1,0){17.5}}
\put(37,455){\line(1,0){18}}
\put(67,450){\circle{25}}
\put(56,446){\line(1,1){15}}
\put(59,441){\line(1,1){17}}
\put(64.5,438.5){\line(1,1){14}}
\put(68,463){\vector(1,1){20}}
\put(68,438){\vector(1,-1){20}}
\put(91,480){5}
\put(91,413){6}
\put(87,450){\circle{16}}
\put(95,450){\vector(3,2){17}}
\put(95,450){\vector(3,-2){17}}
\put(72,461){\vector(1,0){40}}
\put(72,439){\vector(1,0){40}}
\put(120,460){\circle{16}}
\put(120,440){\circle{16}}
\put(128,461){\vector(3,1){20}}
\put(128,461){\vector(3,-1){20}}
\put(128,439){\vector(3,1){20}}
\put(128,439){\vector(3,-1){20}}
\put(151,470){1}
\put(151,453){2}
\put(151,441){3}
\put(151,424){4}
\put(95,466){1}
\put(95,454){\small 2}
\put(95,440){\small 3}
\put(95,428){4}
\put(80,447){\small $0^{uc}$}
\put(113,457){\small $0^{uc}$}
\put(113,437){\small $0^{uc}$}
\put(166,448){$+$}
\put(183,448){4}
\put(200,445){\line(1,0){18}}
\put(200,447){\line(1,0){17.5}}
\put(200,449){\line(1,0){17}}
\put(200,451){\line(1,0){17}}
\put(200,453){\line(1,0){17.5}}
\put(200,455){\line(1,0){18}}
\put(230,450){\circle{25}}
\put(219,446){\line(1,1){15}}
\put(222,441){\line(1,1){17}}
\put(227.5,438.5){\line(1,1){14}}
\put(246,462){\circle{16}}
\put(251.5,468.5){\vector(1,1){15}}
\put(251.5,468.5){\vector(1,-1){18}}
\put(242,450){\vector(1,0){28}}
\put(278,450){\circle{16}}
\put(286,450){\vector(3,1){22}}
\put(286,450){\vector(3,-1){22}}
\put(262,461){1}
\put(259,439){2}
\put(302,462){1}
\put(302,430){2}
\put(268,485){5}
\put(246,437){\circle{16}}
\put(253,432){\vector(3,-1){20}}
\put(253,432){\vector(2,-3){12}}
\put(231,438){\vector(1,-3){8}}
\put(276,422){3}
\put(267,407){4}
\put(241,407){6}
\put(239,459){\small $1^{uu}$}
\put(271,447){\small $0^{uc}$}
\put(239,434){\small $0^{uc}$}
\put(328,448){$+$}
\put(348,448){4}
\put(368,445){\line(1,0){18}}
\put(368,447){\line(1,0){17.5}}
\put(368,449){\line(1,0){17}}
\put(368,451){\line(1,0){17}}
\put(368,453){\line(1,0){17.5}}
\put(368,455){\line(1,0){18}}
\put(398,450){\circle{25}}
\put(387,446){\line(1,1){15}}
\put(390,441){\line(1,1){17}}
\put(395.5,438.5){\line(1,1){14}}
\put(414,462){\circle{16}}
\put(419.5,468.5){\vector(1,1){15}}
\put(419.5,468.5){\vector(1,-1){18}}
\put(410,450){\vector(1,0){28}}
\put(446,450){\circle{16}}
\put(454,450){\vector(3,1){22}}
\put(454,450){\vector(3,-1){22}}
\put(430,461){1}
\put(427,439){2}
\put(470,462){1}
\put(470,430){2}
\put(436,485){5}
\put(414,437){\circle{16}}
\put(421,432){\vector(3,-1){20}}
\put(421,432){\vector(2,-3){12}}
\put(399,438){\vector(1,-3){8}}
\put(444,422){3}
\put(435,407){4}
\put(409,407){6}
\put(407,459){\small $0^{uc}$}
\put(439,447){\small $0^{uc}$}
\put(407,434){\small $0^{uc}$}
\put(10,348){$+$}
\put(24,348){4}
\put(37,345){\line(1,0){18}}
\put(37,347){\line(1,0){17.5}}
\put(37,349){\line(1,0){17}}
\put(37,351){\line(1,0){17}}
\put(37,353){\line(1,0){17.5}}
\put(37,355){\line(1,0){18}}
\put(67,350){\circle{25}}
\put(56,346){\line(1,1){15}}
\put(59,341){\line(1,1){17}}
\put(64.5,338.5){\line(1,1){14}}
\put(84.5,360){\circle{16}}
\put(84.5,340){\circle{16}}
\put(89,367){\vector(1,1){18}}
\put(89,333){\vector(1,-1){18}}
\put(89,367){\vector(1,-1){17}}
\put(89,333){\vector(1,1){17}}
\put(114,350){\circle{16}}
\put(122,350){\vector(1,1){18}}
\put(122,350){\vector(1,-1){18}}
\put(67,330){\circle{16}}
\put(67,322){\vector(1,-2){9}}
\put(67,322){\vector(-1,-2){9}}
\put(98,362){1}
\put(98,331){2}
\put(143,365){1}
\put(143,327){2}
\put(111,385){5}
\put(111,310){6}
\put(79,305){3}
\put(49,305){4}
\put(77.5,357){\small $1^{uu}$}
\put(77.5,337){\small $0^{uc}$}
\put(107,347){\small $0^{uc}$}
\put(60,327){\small $0^{uc}$}
\put(160,348){$+$}
\put(178,348){2}
\put(196,345){\line(1,0){18}}
\put(196,347){\line(1,0){17.5}}
\put(196,349){\line(1,0){17}}
\put(196,351){\line(1,0){17}}
\put(196,353){\line(1,0){17.5}}
\put(196,355){\line(1,0){18}}
\put(226,350){\circle{25}}
\put(215,346){\line(1,1){15}}
\put(218,341){\line(1,1){17}}
\put(223.5,338.5){\line(1,1){14}}
\put(235,368){\circle{16}}
\put(235,332){\circle{16}}
\put(238,353){\vector(3,1){28}}
\put(238,347){\vector(3,-1){28}}
\put(243,370){\vector(3,2){21}}
\put(243,370){\vector(3,-1){23}}
\put(243,330){\vector(3,1){23}}
\put(243,330){\vector(3,-2){21}}
\put(274,360){\circle{16}}
\put(274,340){\circle{16}}
\put(283,360){\vector(3,2){21}}
\put(283,360){\vector(3,-1){23}}
\put(283,340){\vector(3,1){23}}
\put(283,340){\vector(3,-2){21}}
\put(310,371){1}
\put(310,352){2}
\put(311,342){3}
\put(310,319){4}
\put(254,386){5}
\put(254,307){6}
\put(259,368){1}
\put(259,351){2}
\put(259,342){3}
\put(259,325){4}
\put(228,365){\small $1^{uu}$}
\put(228,329){\small $1^{uu}$}
\put(267,357){\small $0^{uc}$}
\put(267,337){\small $0^{uc}$}
\put(327,348){$+$}
\put(345,348){4}
\put(363,345){\line(1,0){18}}
\put(363,347){\line(1,0){17.5}}
\put(363,349){\line(1,0){17}}
\put(363,351){\line(1,0){17}}
\put(363,353){\line(1,0){17.5}}
\put(363,355){\line(1,0){18}}
\put(393,350){\circle{25}}
\put(382,346){\line(1,1){15}}
\put(385,341){\line(1,1){17}}
\put(390.5,338.5){\line(1,1){14}}
\put(402,368){\circle{16}}
\put(402,332){\circle{16}}
\put(405,353){\vector(3,1){28}}
\put(405,347){\vector(3,-1){28}}
\put(410,370){\vector(3,2){21}}
\put(410,370){\vector(3,-1){23}}
\put(410,330){\vector(3,1){23}}
\put(410,330){\vector(3,-2){21}}
\put(441,360){\circle{16}}
\put(441,340){\circle{16}}
\put(450,360){\vector(3,2){21}}
\put(450,360){\vector(3,-1){23}}
\put(450,340){\vector(3,1){23}}
\put(450,340){\vector(3,-2){21}}
\put(477,371){1}
\put(477,352){2}
\put(478,342){3}
\put(477,319){4}
\put(421,386){5}
\put(421,307){6}
\put(426,368){1}
\put(426,351){2}
\put(426,342){3}
\put(426,325){4}
\put(395,365){\small $1^{uu}$}
\put(395,329){\small $0^{uc}$}
\put(434,357){\small $0^{uc}$}
\put(434,337){\small $0^{uc}$}
\put(10,248){$+$}
\put(24,248){2}
\put(37,245){\line(1,0){18}}
\put(37,247){\line(1,0){17.5}}
\put(37,249){\line(1,0){17}}
\put(37,251){\line(1,0){17}}
\put(37,253){\line(1,0){17.5}}
\put(37,255){\line(1,0){18}}
\put(67,250){\circle{25}}
\put(56,246){\line(1,1){15}}
\put(59,241){\line(1,1){17}}
\put(65.5,238.5){\line(1,1){14}}
\put(76,268){\circle{16}}
\put(76,232){\circle{16}}
\put(79,253){\vector(3,1){28}}
\put(79,247){\vector(3,-1){28}}
\put(84,270){\vector(3,2){21}}
\put(84,270){\vector(3,-1){23}}
\put(84,230){\vector(3,1){23}}
\put(84,230){\vector(3,-2){21}}
\put(115,260){\circle{16}}
\put(115,240){\circle{16}}
\put(124,260){\vector(3,2){21}}
\put(124,260){\vector(3,-1){23}}
\put(124,240){\vector(3,1){23}}
\put(124,240){\vector(3,-2){21}}
\put(151,271){1}
\put(152,252){2}
\put(152,242){3}
\put(151,219){4}
\put(95,286){5}
\put(95,207){6}
\put(100,268){1}
\put(100,251){2}
\put(100,242){3}
\put(100,225){4}
\put(69,265){\small $0^{uc}$}
\put(69,229){\small $0^{uc}$}
\put(108,257){\small $0^{uc}$}
\put(108,237){\small $0^{uc}$}
\put(175,248){$+$}
\put(190,248){4}
\put(205,245){\line(1,0){18}}
\put(205,247){\line(1,0){17.5}}
\put(205,249){\line(1,0){17}}
\put(205,251){\line(1,0){17}}
\put(205,253){\line(1,0){17.5}}
\put(205,255){\line(1,0){18}}
\put(235,250){\circle{25}}
\put(224,246){\line(1,1){15}}
\put(227,241){\line(1,1){17}}
\put(232.5,238.5){\line(1,1){14}}
\put(254,257){\circle{16}}
\put(251,237){\circle{16}}
\put(261,261){\vector(1,1){15}}
\put(236,262.5){\vector(3,1){40}}
\put(284,277){\circle{16}}
\put(261,261){\vector(1,-1){12}}
\put(259,235){\vector(1,1){14}}
\put(281,250){\circle{16}}
\put(289,250){\vector(3,2){18}}
\put(289,250){\vector(3,-2){18}}
\put(292,278){\vector(3,2){18}}
\put(292,278){\vector(3,-2){18}}
\put(259,235){\vector(1,-1){16}}
\put(236,238){\vector(1,-2){11}}
\put(315,287){1}
\put(315,264){2}
\put(311,254){3}
\put(311,232){4}
\put(265,280){1}
\put(270,260){2}
\put(262,246){3}
\put(265,232){4}
\put(278,211){5}
\put(250,209){6}
\put(247,254){\small $1^{uu}$}
\put(244,234){\small $0^{uc}$}
\put(277,274){\small $0^{uc}$}
\put(274,247){\small $0^{uc}$}
\put(337,248){$+$}
\put(355,248){4}
\put(370,245){\line(1,0){18}}
\put(370,247){\line(1,0){17.5}}
\put(370,249){\line(1,0){17}}
\put(370,251){\line(1,0){17}}
\put(370,253){\line(1,0){17.5}}
\put(370,255){\line(1,0){18}}
\put(400,250){\circle{25}}
\put(389,246){\line(1,1){15}}
\put(392,241){\line(1,1){17}}
\put(397.5,238.5){\line(1,1){14}}
\put(419,257){\circle{16}}
\put(416,237){\circle{16}}
\put(426,261){\vector(1,1){15}}
\put(401,262.5){\vector(3,1){40}}
\put(449,277){\circle{16}}
\put(426,261){\vector(1,-1){12}}
\put(424,235){\vector(1,1){14}}
\put(446,250){\circle{16}}
\put(454,250){\vector(3,2){18}}
\put(454,250){\vector(3,-2){18}}
\put(457,278){\vector(3,2){18}}
\put(457,278){\vector(3,-2){18}}
\put(424,235){\vector(1,-1){16}}
\put(401,238){\vector(1,-2){11}}
\put(480,287){1}
\put(480,264){2}
\put(476,254){3}
\put(476,232){4}
\put(430,280){1}
\put(435,260){2}
\put(427,246){3}
\put(430,232){4}
\put(443,211){5}
\put(415,209){6}
\put(412,254){\small $0^{uc}$}
\put(409,234){\small $1^{uu}$}
\put(442,274){\small $0^{uc}$}
\put(439,247){\small $0^{uc}$}
\put(10,148){$+$}
\put(28,148){4}
\put(43,145){\line(1,0){18}}
\put(43,147){\line(1,0){17.5}}
\put(43,149){\line(1,0){17}}
\put(43,151){\line(1,0){17}}
\put(43,153){\line(1,0){17.5}}
\put(43,155){\line(1,0){18}}
\put(73,150){\circle{25}}
\put(62,146){\line(1,1){15}}
\put(65,141){\line(1,1){17}}
\put(70.5,138.5){\line(1,1){14}}
\put(92,157){\circle{16}}
\put(89,137){\circle{16}}
\put(99,161){\vector(1,1){15}}
\put(74,162.5){\vector(3,1){40}}
\put(122,177){\circle{16}}
\put(99,161){\vector(1,-1){12}}
\put(97,135){\vector(1,1){14}}
\put(119,150){\circle{16}}
\put(127,150){\vector(3,2){18}}
\put(127,150){\vector(3,-2){18}}
\put(130,178){\vector(3,2){18}}
\put(130,178){\vector(3,-2){18}}
\put(97,135){\vector(1,-1){16}}
\put(74,138){\vector(1,-2){11}}
\put(153,187){1}
\put(153,164){2}
\put(149,154){3}
\put(149,132){4}
\put(103,180){1}
\put(108,160){2}
\put(100,146){3}
\put(103,132){4}
\put(116,111){5}
\put(88,109){6}
\put(85,154){\small $0^{uc}$}
\put(82,134){\small $0^{uc}$}
\put(115,174){\small $0^{uc}$}
\put(112,147){\small $0^{uc}$}
\put(175,148){$+$}
\put(193,148){2}
\put(210,145){\line(1,0){18}}
\put(210,147){\line(1,0){17.5}}
\put(210,149){\line(1,0){17}}
\put(210,151){\line(1,0){17}}
\put(210,153){\line(1,0){17.5}}
\put(210,155){\line(1,0){18}}
\put(240,150){\circle{25}}
\put(229,146){\line(1,1){15}}
\put(232,141){\line(1,1){17}}
\put(237.5,138.5){\line(1,1){14}}
\put(249,168){\circle{16}}
\put(249,132){\circle{16}}
\put(260,150){\circle{16}}
\put(268,150){\vector(1,1){12}}
\put(268,150){\vector(1,-1){12}}
\put(257,170){\vector(3,2){21}}
\put(257,170){\vector(3,-1){23}}
\put(257,130){\vector(3,1){23}}
\put(257,130){\vector(3,-2){21}}
\put(288,160){\circle{16}}
\put(288,140){\circle{16}}
\put(297,160){\vector(3,2){21}}
\put(297,160){\vector(3,-1){23}}
\put(297,140){\vector(3,1){23}}
\put(297,140){\vector(3,-2){21}}
\put(324,171){1}
\put(325,152){2}
\put(325,142){3}
\put(324,119){4}
\put(268,186){5}
\put(268,107){6}
\put(273,168){1}
\put(267,156){2}
\put(267,137){3}
\put(273,125){4}
\put(242,165){\small $1^{uu}$}
\put(242,129){\small $1^{uu}$}
\put(253,147){\small $1^{cc}$}
\put(281,157){\small $0^{uc}$}
\put(281,137){\small $0^{uc}$}
\put(0,70){Fig. 1. Graphical representation of the amplitude $A_2^{0^{uc}0^{uc}}$ for the
case of $\Sigma_c\Sigma_c$, $\Sigma^*_c\Sigma^*_c$, $I=2$, $J^P=0^+$.}
\end{picture}

\end{document}